\RequirePackage{lineno}
\documentclass[manuscript]{aastex63}

\submitjournal{ApJS}

\shorttitle{}
\shortauthors{Die Duan et al.}
\graphicspath{{./}{figures/}}
\begin{document}

\title{The Radial Dependence of Proton-scale Magnetic Spectral Break in Slow Solar Wind during \textit{PSP} Encounter 2}

\correspondingauthor{Jiansen He}
\email{jshept@pku.edu.cn}

\correspondingauthor{Die Duan}
\email{dduan@pku.edu.cn}

\correspondingauthor{Trevor A. Bowen}
\email{tbowen@berkeley.edu}

\author{Die Duan}
\affil{School of Earth and Space Sciences, Peking University, Beijing, 100871, China}
\affil{Space Sciences Laboratory, University of California, Berkeley, CA 94720-7450, USA}

\author[0000-0002-4625-3332]{Trevor A. Bowen}
\affil{Space Sciences Laboratory, University of California, Berkeley, CA 94720-7450, USA}

\author{Christopher H. K. Chen}
\affil{School of Physics and Astronomy, Queen Mary University of London, London E1 4NS, UK}

\author{Alfred Mallet}
\affil{Space Sciences Laboratory, University of California, Berkeley, CA 94720-7450, USA}

\author[0000-0001-8179-417X]{Jiansen He}
\affil{School of Earth and Space Sciences, Peking University, Beijing, 100871, China}

\author[0000-0002-1989-3596]{Stuart D. Bale}
\affiliation{Space Sciences Laboratory, University of California, Berkeley, CA 94720-7450, USA}
\affiliation{School of Physics and Astronomy, Queen Mary University of London, London E1 4NS, UK}
\affiliation{Physics Department, University of California, Berkeley, CA 94720-7300, USA}

\affiliation{The Blackett Laboratory, Imperial College London, London, SW7 2AZ, UK}

\author{Daniel Vech}
\affiliation{Climate and Space Sciences and Engineering, University of Michigan, Ann Arbor, MI 48109, USA}
\affiliation{Laboratory for Atmospheric and Space Physics, University of Colorado, Boulder, CO 80303, USA}

\author[0000-0002-7077-930X]{J. C. Kasper}
\affiliation{Climate and Space Sciences and Engineering, University of Michigan, Ann Arbor, MI 48109, USA}
\affiliation{Smithsonian Astrophysical Observatory, Cambridge, MA 02138 USA}

\author[0000-0002-1573-7457]{Marc Pulupa}
\affiliation{Space Sciences Laboratory, University of California, Berkeley, CA 94720-7450, USA}

\author[0000-0002-0675-7907]{John W. Bonnell}
\affil{Space Sciences Laboratory, University of California, Berkeley, CA 94720-7450, USA}

\author[0000-0002-3520-4041]{Anthony W. Case}
\affiliation{Smithsonian Astrophysical Observatory, Cambridge, MA 02138 USA}

\author[0000-0002-4401-0943]{Thierry {Dudok de Wit}}
\affiliation{LPC2E, CNRS and University of Orl\'eans, Orl\'eans, France}

\author[0000-0003-0420-3633]{Keith Goetz}
\affiliation{School of Physics and Astronomy, University of Minnesota, Minneapolis, MN 55455, USA}

\author[0000-0002-6938-0166]{Peter R. Harvey}
\affiliation{Space Sciences Laboratory, University of California, Berkeley, CA 94720-7450, USA}

\author[0000-0001-6095-2490]{Kelly E. Korreck}
\affiliation{Smithsonian Astrophysical Observatory, Cambridge, MA 02138 USA}

\author{Davin Larson}
\affiliation{Space Sciences Laboratory, University of California, Berkeley, CA 94720-7450, USA}

\author{Roberto Livi}
\affiliation{Space Sciences Laboratory, University of California, Berkeley, CA 94720-7450, USA}

\author[0000-0003-3112-4201]{Robert J. MacDowall}
\affiliation{Solar System Exploration Division, NASA/Goddard Space Flight Center, Greenbelt, MD, 20771}

\author[0000-0003-1191-1558]{David M. Malaspina}
\affiliation{Laboratory for Atmospheric and Space Physics, University of Colorado, Boulder, CO 80303, USA}

\author[0000-0002-7728-0085]{Michael Stevens}
\affiliation{Smithsonian Astrophysical Observatory, Cambridge, MA 02138 USA}

\author[0000-0002-7287-5098]{Phyllis Whittlesey}
\affiliation{Space Sciences Laboratory, University of California, Berkeley, CA 94720-7450, USA}



\begin{abstract}

Magnetic field fluctuations in the solar wind are commonly observed to follow a power law spectrum. Near proton-kinetic scales, a spectral break occurs which is commonly interpreted as a transition to kinetic turbulence. However, this transition is not yet entirely understood. By studying the scaling of the break with various plasma properties, it may be possible to constrain the processes leading to the onset of kinetic turbulence. Using data from Parker Solar Probe (\textit{PSP}), we measure the proton scale break over a range of heliocentric distances, enabling a measurement of the transition from inertial to kinetic scale turbulence under various plasma conditions. We find that the break frequency $f_b$ increases as the heliocentric distance $r$ decreases in the slow solar wind following a power law $f_b\sim r^{-1.11}$. We also compare this to the characteristic plasma ion scales to relate the break to the possible physical mechanisms occurring at this scale. The ratio between $f_b$ and $f_c$, the Doppler shifted ion cyclotron resonance scale, is approximately unity for all plasma $\beta_p$. At high $\beta_p$ the ratio between $f_b$ and $f_\rho$, the Doppler shifted gyroscale, is approximately unity; while at low $\beta_p$ the ratio between $f_b$ and $f_d$, the Doppler shifted proton-inertial length is unity. Due to the large comparable Alfv\'en and solar wind speeds, we analyze these results using both the standard and modified Taylor hypothesis, demonstrating robust statistical results.
%, and find the break scale is closest to the proton cyclotron resonance scale. These results suggest the physical processes at the proton cyclotron resonance scale play an important role in the development of the slow solar wind turbulence in the inner heliosphere.

\end{abstract}



\section{Introduction} 
\label{sec:intro}

Understanding kinetic dissipation in magnetized plasma is essential for explaining the physical origin and evolution of the solar wind. Observationally, the power spectral density (PSD) of the magnetic field fluctuations is commonly divided into two regimes separated by a spectral break. The lower frequencies, corresponding to larger physical scales, correspond to magnetohydrodynamic (MHD) fluctuations, with an inertial range of turbulence similar to Kolmogorov $f^{-5/3}$ power-law spectrum. In the high frequency range, the power spectra is observed to steepen with a spectral index of between -2 to -4 \citep{Bruno2013, Kiyani2015, Chen2016}. These scales are thought to correspond to scales, in which the MHD approximation is no longer valid, and kinetic effects of the protons should be considered \citep{Alexandrova2009}. However, the specific processes occurring in the kinetic range have not been determined, with significant debate regarding the nature of the fluctuations and the relevant non-linear processes \citep{Howes2017Ph}.

The steepening of the spectral index possibly implies that cascaded energy at the end of MHD scale may be gradually dissipated or develop into a dispersive kinetic turbulence. Observationally, the solar wind expands non-adiabatically indicating that {\em{in-situ}} heating must occur. Dissipation of the inertial range turbulence is one source of energy capable of proton heating, though there are multiple mechanisms which may lead to dissipation \citep{Marsch2006}. Kinetic Alfv\'en waves (KAW) may dissipate via Landau damping near the scale of the proton gyroradius $\rho_p = v_{th,p}/\Omega_p$, where $v_{th,p}$ is the thermal velocity of proton and $\Omega_p=eB/m_p$ is the proton gyrofrequency, $e$ is the elementary charge, B is the mean magnetic field and $m_p$ is the mass of the proton \citep{Leamon1999a,Schekochihin2009}. Stochastic proton heating is also a possible dissipation mechanism at scales near $\rho_p$. The ions could be heated perpendicularly when the amplitude of the gyro-scale fluctuations is large \citep{Chandran2010,Bourouaine2013,Vech2017,martinovic2019radial}. The proton inertial length $d_p = v_A/\Omega_p$ is another important scale associated with dissipation, where $v_A = B/\sqrt{\mu_0n_pm_p}$ is the Alfv\'en speed, with $\mu_0$ being the vacuum magnetic permeability and $n_p$ being the proton density. The proton inertial length corresponds to the scale at which electrons can decouple from protons and it may limit the size of small-scale current sheets formed through non-linear turbulent processes, which in turn may dissipate energy through magnetic reconnection \citep{Leamon2000,Dmitruk2004,Vasquez2007}. 

Alfv\'en waves with quasi-parallel propagation at relatively higher frequency may dissipate through cyclotron resonance damping. For parallel propagating Alfv\'en waves, the damping will occur at the (parallel) wavenumber corresponding to the cyclotron resonance $k_c=\Omega_p/(v_A+v_{th,p})$ \citep{Leamon1998}. Studies of anisotropy in solar wind turbulence using the method introduced by \citet{Cho2000} and \citet{Horbury2008} suggest that the inertial range is highly anisotropic near the kinetic break with $k_\perp \gg k_\parallel$, such that most the energy is contained in perpendicular fluctuations which do not have parallel wavenumbers resonant with parallel cyclotron waves \citep{Chen2010b}. The 2D PSD distribution ($k_\parallel$, $k_\perp$) as reconstructed with the tomography method based on Fourier projection-slice theorem reveals the dominance of oblique propagation of Alfv\'enic fluctuations extending its power ridge to higher $k_\perp$ and also higher $k_\parallel$, which indicates the existence of oblique Alfven-cyclotron waves \citep{He2013, Yan2016}.  

Alternatively, the change of the spectral slope may indicate a transition from a cascade of non-dispersive Alfv\'en waves to a cascade of dispersive kinetic Alfv\'en waves around $k_\perp \rho_p \sim 1$\citep{Bale2005a,Howes2008,Schekochihin2009}. It has been additionally suggested that a cascade of whistler modes or magnetosonic waves may develop at kinetic scales \citep{Stawicki2001,PeterGary2009}. Furthermore, the inclusion of the Hall term in the MHD approximation has been proposed as the source of the break at scales $d_pk_\perp\sim 1$ \citep{Galtier2006}. \citet{Mallet2017} and \citet{Loureiro2017} suggest that the inertial-range turbulence could generate sheet-like turbulent structures, which could be disrupted by reconnection below a disruption scale intermediate to $d_p$ and $\rho_p$.

Given the number of potential mechanisms which generate a spectral break, and the relatively narrow range in the physical scales predicted, distinguishing these various mechanisms using empirical measurements has proven a difficult task \citep{markovskii2008statistical}. Furthermore, these different physical processes may occur simultaneously in the solar wind, complicating efforts to quantify their relative contributions \citep{verscharen2019multi}.

Many previous studies have explored the transition from inertial to kinetic scale physical processes through both observations and simulations, although no consensus has been reached. Observationally, the mechanisms which lead to spectral steepening may be constrained by investigating the dependence of the spectral break frequency on various plasma parameters. For example, the $\beta_p$-dependence of the break scale has been studied at 1 AU using \textit{WIND} data, where $\beta_p  = \rho_p^2/d_p^2$ is the ratio of proton thermal pressure to magnetic pressure. For example, \citet{Chen2014} found the break frequency ($f_b$) close to $f_{d}$ at $\beta_p \ll 1$ and close to $f_{\rho}$ at $\beta_p\gg 1$, where $f_{d}=v_{sw}/(2\pi d_p)$ and $f_{\rho} = v_{sw}/(2\pi \rho_p)$ are the frequencies corresponding to the spatial scales $d_p$ and $\rho_p$ in the spacecraft frame under the Taylor Hypothesis, which approximates the observed time evolution of fluctuations in the spacecraft frame as spatial structures advected at the solar wind speed $v_{sw}$ . Numerical 2D-hybrid simulations found similar $\beta_p$ dependence \citep{Franci2016}. \citet{Wang2018} found $f_b/f_{d}$ is statistically independent with $\beta_{p}$ of $0.1 < \beta_{p} < 1.3$ plasma. \citet{Woodham2018} and \citet{Duan2018} suggest that the spectral break is best associated with the proton cyclotron scale $f_c=v_{sw}k_c/(2\pi)$. \citet{Vech2018} proposed the break may be caused by magnetic reconnection at a disruption scale intermediate to $d_p$ and $\rho_p$ predicted in \citet{Mallet2017}. The spectral break is found to be independent of $\theta_{\rm{VB}}$, the angle between solar wind velocity and magnetic field, indicating that the spectral break seems to be isotropic in the wavenumber space \citep{Duan2018}. \citet{Duan2018} further proposed and illustrated that the breakdown of magnetic frozen-in condition in wavenumber space, as a combination of dissipation and dispersion effects, could be a more isotropic explanation as compared to the dissipation or the dispersion alone.

Several studies investigated the break scale at different heliocentric distances and its relation with plasma scales. \citet{Perri2010} suggested the break frequency did not show any remarkable evolution between 0.3 AU and 4.9 AU based on observations from \textit{MESSENGER} and \textit{Ulysess}. \citet{Bourouaine2012} also found the break frequency $f_b$ does not change significantly from 0.3 to 0.9 AU from \textit{Helios 2}, and $f_b$ follows $f_d$ if assuming a 2D turbulence model. \citet{Bruno2014} found the break moves to higher frequencies as the heliocentric distance decreases, finding agreement with the proton cyclotron resonance scale between 0.42 and 5.3 AU. While many previous studies have focused on the radial behavior of the spectral break in the fast solar wind, the scaling of the spectral break in the slow wind has not been investigated.

NASA's Parker Solar Probe (\textit{PSP}) provides a set of {\em{in-situ}} instruments capable of constraining the kinetic processes which contribute to heating and acceleration in the corona and nascent solar wind \citep{Fox2016,Bale2016,Kasper2016}. This manuscript provides a statistical analysis of the behavior of the proton-scale spectral break observed by \textit{PSP} between 0.17 AU and 0.63 AU, and its radial dependence in the slow solar wind. By measuring the radial dependence of the break we are able to compare the location of the spectral break with various physical scales under a range of plasma conditions, enabling an investigation into the mechanisms behind spectral steepening of the kinetic range.

\section{Data and Method}
\label{sec:data}
\begin{figure}[ht!]
\plotone{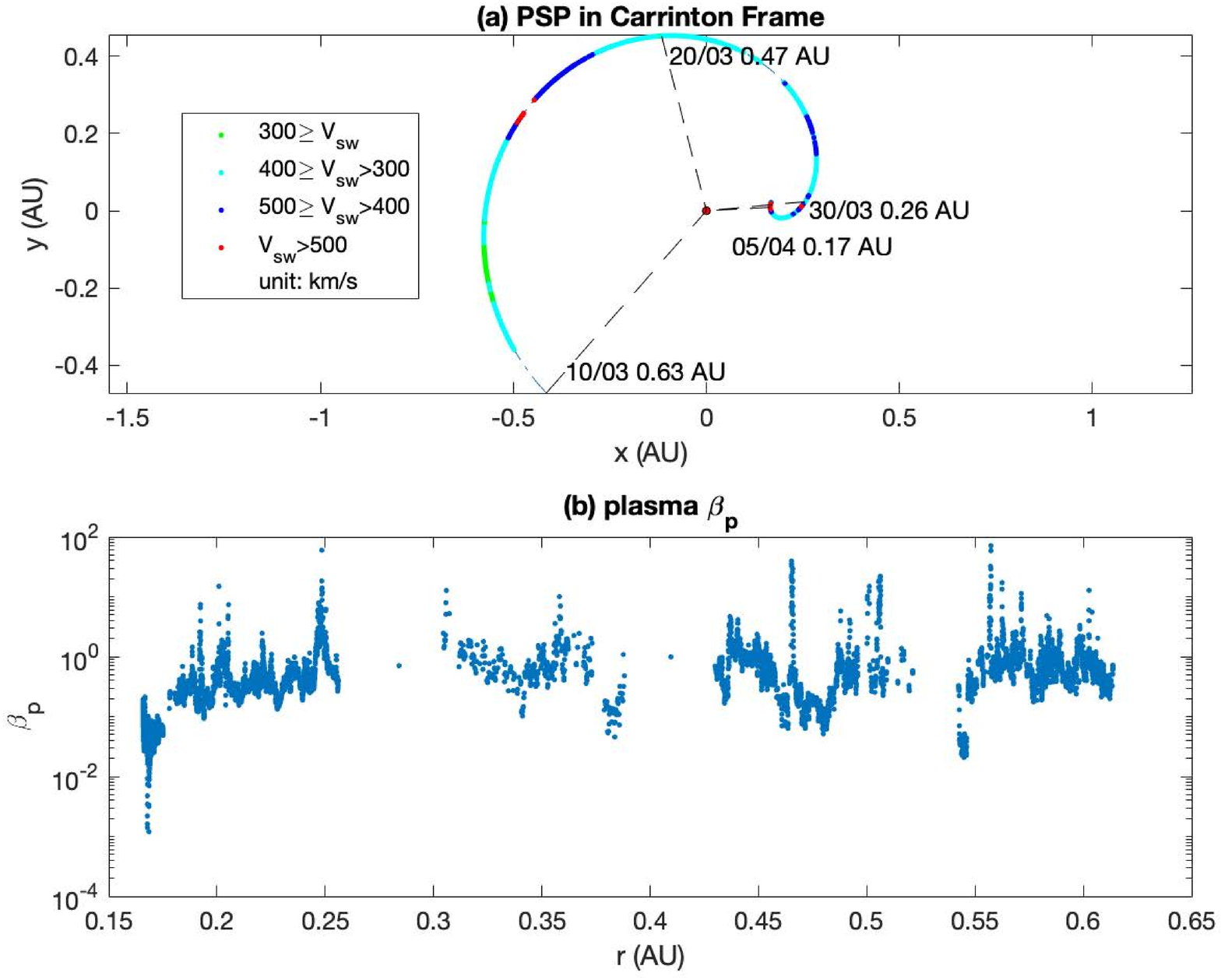}
\caption{(a) The location of \textit{PSP} during Encounter 2 in the corotating Carrington frame. The red solid circle at the origin is the Sun. The heliocentric distance of \textit{PSP} is decreasing. Black dashed lines indicate the location of \textit{PSP}  at several times. The orbit color is used to indicate different solar wind speeds. The thin dashed line on the orbit means the SPC data was unavailable. (b) $\beta_p$ at different distances. Blank regions indicate unavailable data. \label{fig:1}}
\end{figure}

We analyze 26 days of data from \textit{PSP} during the cruise phase of the second orbit of \textit{PSP} from Mar 10, 2019 to Apr 5, 2019, data on Mar 16 were excluded as the time resolution of the magnetic field is not sufficient to resolve the spectral break. During the period, \textit{PSP} covers the distance between 0.63 AU (Mar 10) and 0.17 AU (Apr 5) from the Sun. Magnetic field measurements on \textit{PSP} are made by the FIELDS/MAG fluxgate magnetometer \citep{Bale2016}. Measurements of the solar wind speed, thermal speed and proton density by the SWEAP/SPC instrument are used to compute plasma scales \citep{Kasper2016}. Sample rates of FIELDS and SWEAP data vary between the different mission phases and encounters. Between March, 10 2019 and March 31, 2019, \textit{PSP} was in cruise phase with low cadence (MAG 9.2 Hz, SPC 0.036 Hz) sample rate. From March 31, 2019 to April 4, 2019 the mission was in encounter phase near perihelion, and higher cadence measurements are obtained (MAG 149 Hz, SPC 5 Hz). Figure \ref{fig:1} (a) shows an overview of the trajectory of the \textit{PSP} in the rotating Carrington heliographic frame. For the majority of the orbit \textit{PSP} is in slow solar wind ($v_{SW}<$ 500 km/s). There are no intervals with average $v_{SW}>$ 500 km/s. Figure \ref{fig:1} (b) shows $\beta_p$ as a function of the heliocentric distance $r$. As the distance between \textit{PSP} and the Sun decreases, the proton plasma $\beta$ also decreases due to the increasing strength of the magnetic field; typically, $\beta_p<1$.

The trace power spectral density is estimated by applying a continuous moving window transform on the vector magnetic field. The 26 day interval is divided into partially overlapping 10 minute segments. The beginnings of each adjacent segments are 2.5 minutes apart (overlapping 75\%). A Hanning window is used to reduced spectral leakage in each segment. For each segment the power spectrum is taken using an ensemble average of five adjacent segments. Each PSD actually correspond to data of 20 minutes.

To locate the proton-scale spectral break, we employed the method of \citet{Bruno2014} and \citet{Wang2018}. Two frequency ranges at either end of the spectrum are a priori selected as the  inertial (between 0.1 Hz and 0.5 Hz) and dissipation ranges. Table \ref{tab:1} highlights the range of frequencies for the dissipation spectra over the orbit. A least-squares linear fit of a power law in logarithmic space is performed on the data over each range. The break frequency $f_b$ is defined as the intersection of the two fitting lines.  Because the range of spacecraft frequencies corresponding to the dissipation range changes with heliocentric distance, the range over which the fit is performed is varied throughout the orbit.  Additionally, spectral flattening is observed when the amplitude of the turbulent fluctuations reaches the noise level of the MAG ($10^{-3}\sim 10^{-4}\ \rm{nT^2/Hz}$). Because of the decreasing strength of the fluctuations at larger distances, the noise floor is reached at lower frequencies in the cruise data. 

\begin{figure}[ht!]
\plotone{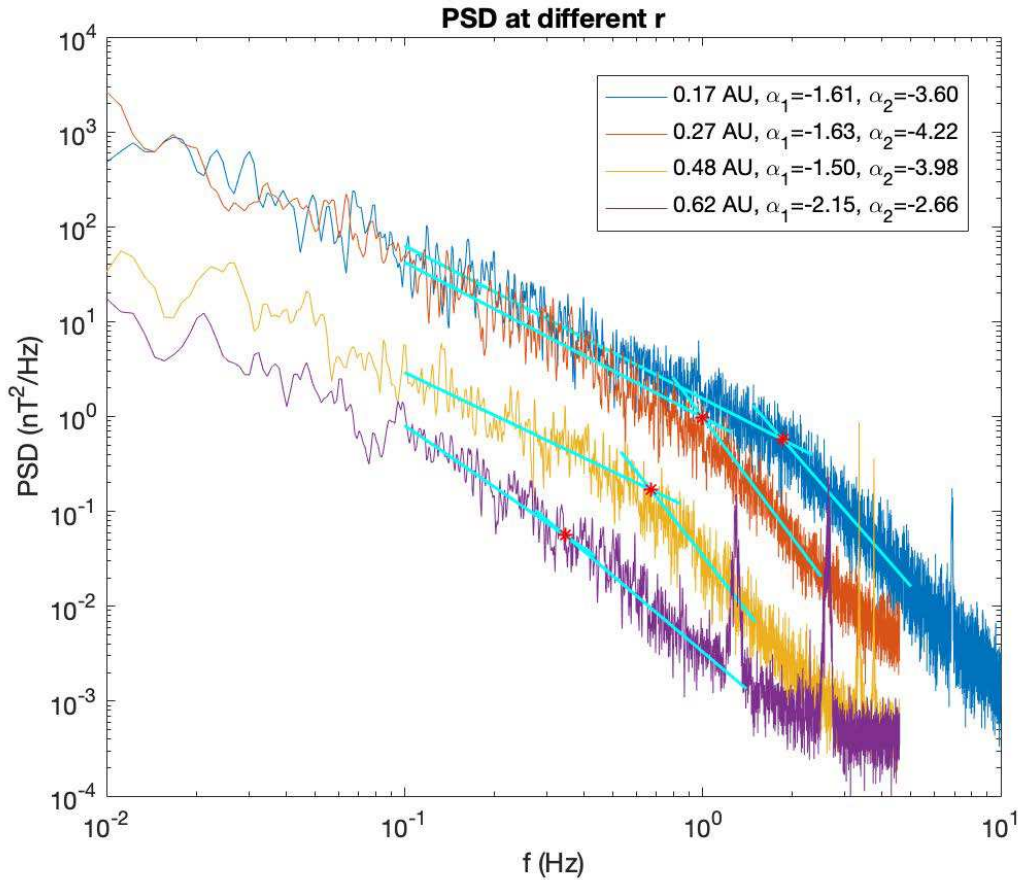}
\caption{Examples of the PSDs of magnetic field fluctuations at several heliocentric distances. The cyan lines indicate fitted power law spectra. The red stars are intersections of the fitted lines and defined as the break frequency $f_b$. The fitted inertial range index $\alpha_1$ and fitted dissipation range index $\alpha_2$ are shown in the legend. \label{fig:2}}
\end{figure}

Figure \ref{fig:2} shows an example of power density spectra at several distances with measured spectral indices and breaks. At larger distances, the spectral break shifts to lower spacecraft frame frequency. The top three PSDs shows a typical inertial range slope $-5/3<\alpha_1<-3/2$, and a dissipation range slope $\alpha_2\approx -4$.  The spectra from 0.62 AU does not show an obvious break between two power law spectra. Additionally the inertial range spectral index is somewhat steeper than what is typically observed. This shape has been previously reported by \citet{Bruno2014b} in slow winds. \citet{Bowen2018} demonstrates that the presence of steep magnetic spectra (i.e. $\alpha_1 \sim -2$) likely corresponds to observations of intermittency in the turbulent fluctuations.

We removed several intervals with spectral features peaked at ion scales, which results in a deviation from power law distributions. The presence of these features is likely a secondary population of ion cyclotron waves \citep{Bowen2019}. To systematically control for effects from secondary population of fluctuations, we only accept spectra that fall within a range of spectral indices statistically consistent with known turbulent scalings  $-2.5<\alpha_1 <-1.2$ and $\alpha_1 > \alpha_2$ In total 14820 intervals were obtained with  10724 of them returning $\alpha_1$ and $\alpha_2$ within our constrained bounds. 5194 of these intervals have corresponding particle data. Mean values of $v_{sw}$, $v_{th,p}$, $n_p$ are averaged over each of intervals. $k_c,d_p,\rho_p,\beta_p$ are calculated from the plasma data. We find that $\beta_p<1$ in 4479 intervals, and $\beta_p>1$ in 715 intervals. 

Under the Taylor Hypothesis, the relation between the wavevector of the fluctuation $\mathbf{k}$ and the corresponding frequency $f$ in the spacecraft frame is $2\pi f = \mathbf{k}\cdot\mathbf{v}_{sw}$. Several possible assumptions can possibly made for simplifying the wavevector direction relative to the solar wind flow. If the fluctuations propagate along the solar wind direction, $2\pi f = k v_{sw}$. If the fluctuations propagate parallel to the mean magnetic field direction, $2\pi f = k v_{sw} \cos(\theta_{VB})$. If quasi-2D turbulence with dominant perpendicular fluctuations is assumed, then $\omega = k_\perp v_{sw} \sin(\theta_{VB})\cos(\phi)$, where $\phi$ is the angle between the wavevector and the ($\mathbf{v}_{sw}$, $\mathbf{B}$) plane \citep{Bourouaine2012}. \citet{Duan2018} found that the spectral break frequency is invariant with the magnetic field's orientation, suggesting that the approximation $ 2\pi f=kv_{sw}$ is appropriate. The corresponding frequencies for the physical scales are $f_c=v_{sw}k_c/(2\pi)$, $f_d=v_{sw}/(2\pi d_p)$, $f_\rho=v_{sw}/(2\pi \rho_p)$.

%$f_c=v_{sw}k_c/(2\pi), f_d=v_{sw}/(2\pi d_p), f_\rho=v_{sw}/(2\pi \rho_p)$. The located break frequency $f_b$ corresponds to the spatial break scale $\mathbf{k}_b$, where $f_b=\mathbf{k}_b\cdot\mathbf{v}_{sw}/(2\pi)$}.

\begin{figure}[ht!]
\plotone{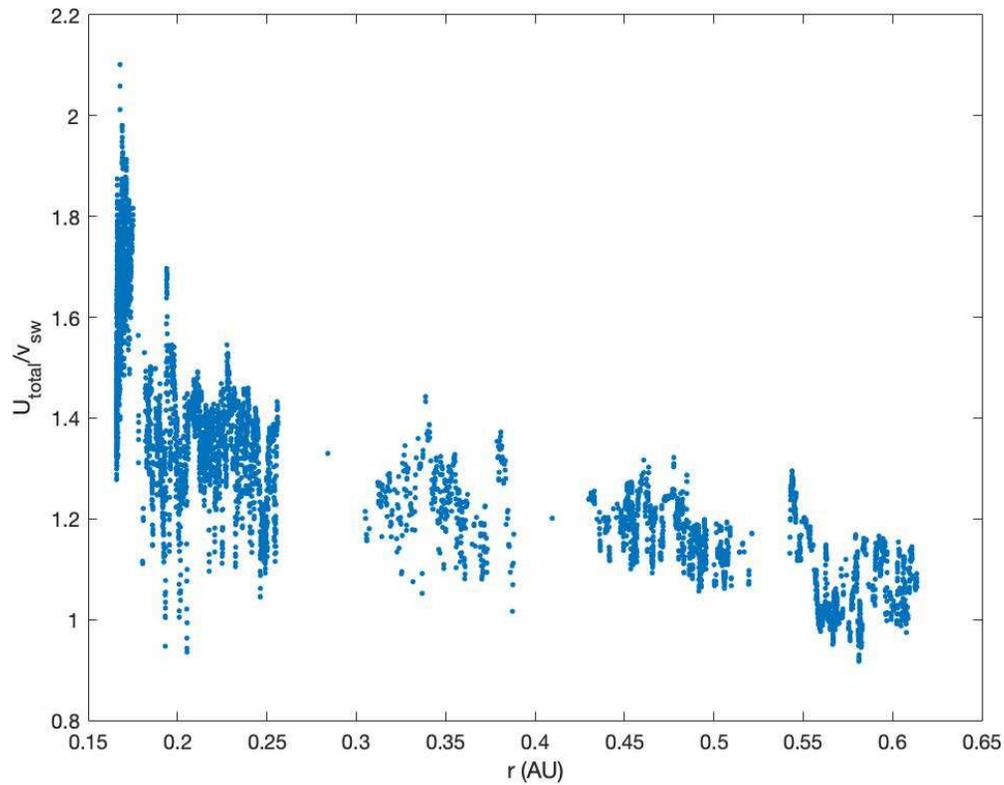}
\caption{The ratio of $U_{total}$ to $v_{sc}$ during the cases. \label{fig:6}}
\end{figure}

Due to the comparable Alfv\'en and solar wind, and spacecraft speeds, it is unclear whether the Taylor hypothesis is valid for \textit{PSP} observations during its perihelion \citep{Narita2013, Bourouaine2018, Bourouaine2019, Chhiber2019}. Recent work from \citet{Chaspis2019} suggests the Taylor hypothesis may not be applicable when \textit{PSP} is below 40 solar radii (0.19 AU). To verify our results against the assumption of the Taylor hypothesis, we apply an analysis of the proton break scaling to the modified Taylor Hypothesis: $2\pi f^* = \mathbf{k}\cdot\mathbf{U}_{total}$ \citep{Klein2015}. Here $\mathbf{U}_{total}=\mathbf{v}_{sw}+\mathbf{v}_A-\mathbf{v}_{sc}$, and $\mathbf{v}_{sc}$ is the velocity of the \textit{PSP}. The modified Taylor hypothesis assumes that the anti-sunward propagating fluctuations are approximately frozen into a frame with velocity $\mathbf{U}_{total}$ if the fluctuations do not grow or damp significantly when passing over the spacecraft. The modified corresponding characteristic frequencies are $f_c^*=U_{total}k_c/(2\pi)$, $f_d^*=U_{total}/(2\pi d_p)$, and $f_\rho^*=U_{total}/(2\pi \rho_p)$, where $U_{total} = |\mathbf{U}_{total}|$. Figure \ref{fig:6} shows $U_{total}/v_{sw}$ during our cases. The ratio is almost greater than 1 (97\% of cases), making the modified characteristic frequencies smaller, especially below 0.19 AU. This modified Taylor hypothesis could hold as the outward-propagating fluctuations are dominant near the perihelion \citep{Chen2019}.

\section{Results}
\label{sec:result}
\begin{figure}[ht!]
\plotone{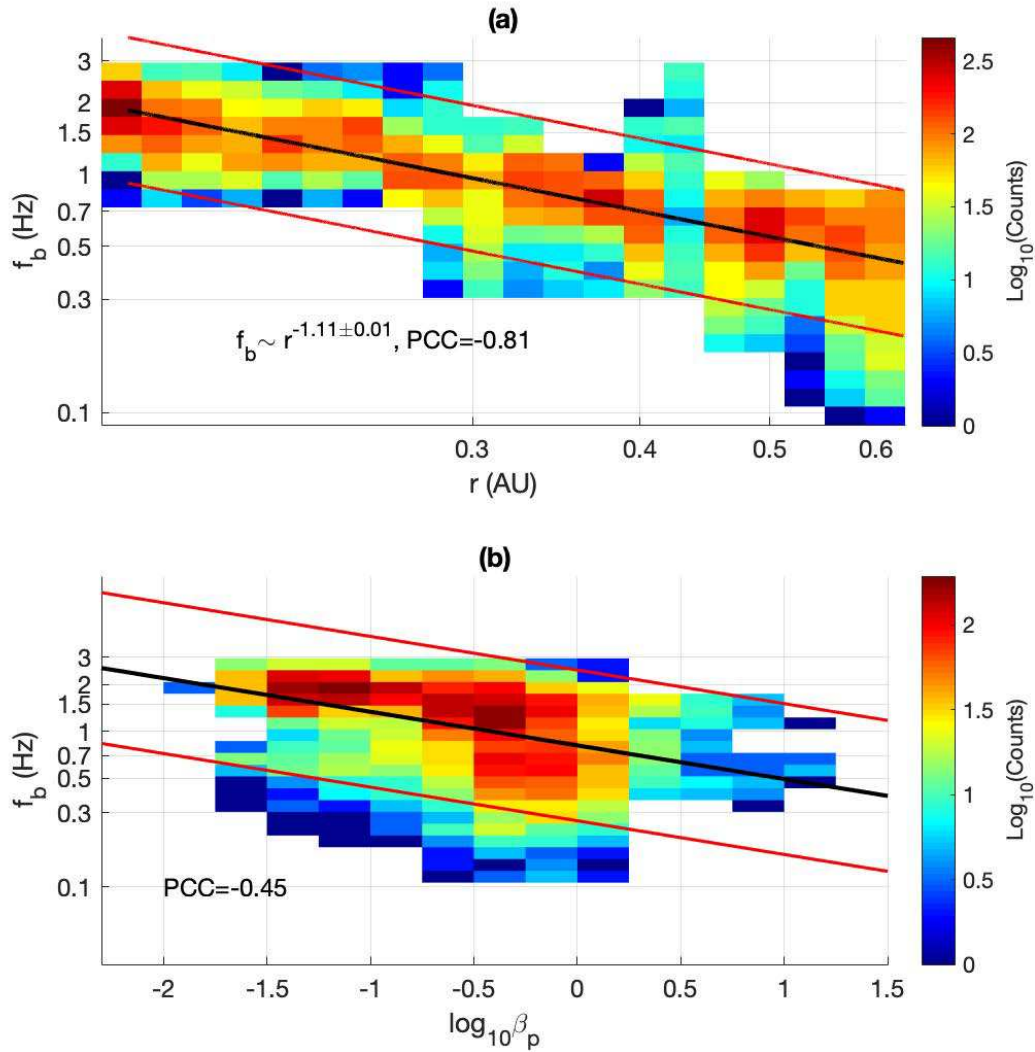}
\caption{(a) 2D-histogram of the measured break frequencies $f_b$ with  heliocentric distance $r$. The black lines are the result of linear regression in log-log space, and red lines indicate the 95\% confidence interval estimated by the standard deviation of the regression. (b) 2D-histogram of the measured break frequencies $f_b$ with $\beta_p$; black and red lines again show the results of linear regression and confidence intervals. Panel(a) contains all 10724 measured intervals while panel(b) only contains intervals with available SPC data. \label{fig:3}}
\end{figure}

Figure \ref{fig:3}(a)  shows the distribution of break frequency $f_b$ with heliocentric distance $r$. Figure \ref{fig:3}(b) shows the  distribution of $f_b$ with $\beta_p$. The data are binned in a 20 $\times$ 20 grid in log-log space. There is large variation in $f_b$ and a clear radial dependence with a power law of $f_b \sim r^{-1.11\pm 0.01}$. A Pearson correlation coefficient is calculated with PCC($r,f_b$) = -0.81, and a Spearman correlation coefficient SCC($r,f_b$) = -0.84. This result is similar to the scalings in the fast solar wind suggested by \citet{Bruno2014}. This radial trend is also consistent with the outer-scale break of the PSD \citep{Chen2019}. 

The $f_b$ shows a weak dependence with $v_{sw}$ with PCC($v_{sw},f_b$) = 0.14 and SCC($v_{sw},f_b$) = 0.10. $f_b$ also decreases with $\beta_p$; PCC($\beta_p,f_b$) = -0.49, SCC($\beta_p,f_b$) = -0.51.

To investigate the correlation between $f_b$ and physical plasma scales, we calculated average $f_{\rho }$, $f_{d}$, and $f_{c}$ for each interval having the measurement of particle data.  Table \ref{tab:1} shows that $f_b$ is correlated with all of these scales to a similar degree. It is accordingly difficult to uniquely distinguish the scale which best represents the break frequency.

\begin{figure}[ht!]
\plotone{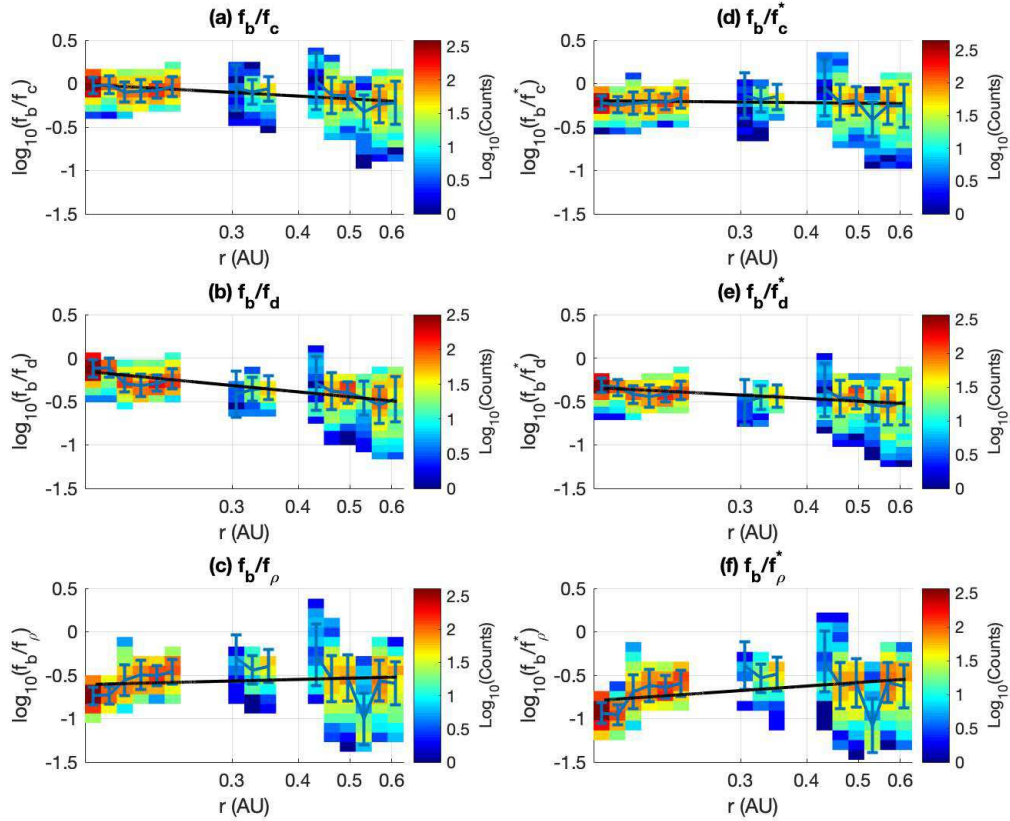}
\caption{The 2D-histogram of the distribution of the occurrence of (a) $\log_{10}(f_b/f_c)$, (b) $\log_{10}(f_b/f_d)$, (c) $\log_{10}(f_b/f_\rho)$, (d) $\log_{10}(f_b/f_c^*)$, (e) $\log_{10}(f_b/f_d^*)$, (f) $\log_{10}(f_b/f_\rho^*)$ over the heliocentric distance $r$ respectively. The starred frequencies are the corresponding frequencies from the modified Taylor hypothesis. The black lines are the linear fitting with least-squares method. The averages and standard deviations of each $r$ bin are plotted as blue lines.\label{fig:4}}
\end{figure}

The ratio of $f_b$ to these characteristic frequencies are calculated and illustrated in Figure \ref{fig:4} (a), (b) and (c). The data is again binned in a 20 $\times$ 20 grid in log-log space. The average and the stand deviation inside each bin are illustrated with blue lines. The average and the standard deviation of each ratio over all of the data is $0.87\pm 0.34 (f_b/f_{c})$, $0.56\pm 0.24 (f_b/f_{d})$ and $0.32\pm 0.22 (f_b/f_{\rho})$. The spectral break occurs nearest the cyclotron resonance frequency. The average $f_b/f_c$ is the largest in each bin. $f_b/f_c$ and $f_b/f_d$ decrease as the distance become larger, while $f_b/f_\rho$ is opposite. Panel (d-f) shows the ratio of modified frequencies. We get the same result assuming the modified Taylor hypothesis. 

\begin{figure}[ht!]
\plotone{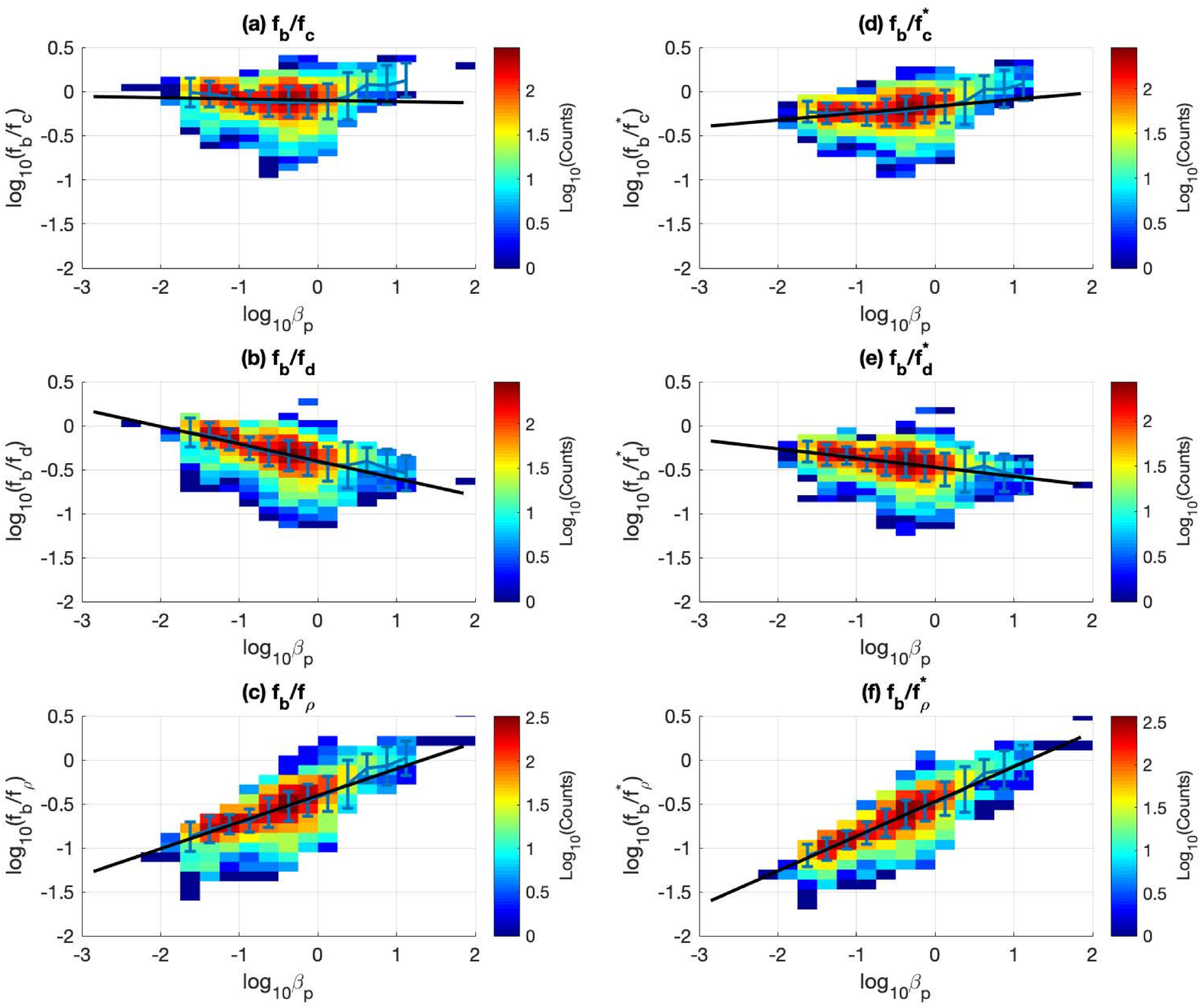}
\caption{The 2D-histogram of the distribution of the occurrence of (a) $\log_{10}(f_b/f_c)$, (b) $\log_{10}(f_b/f_d)$, (c) $\log_{10}(f_b/f_\rho)$, (d) $\log_{10}(f_b/f_c^*)$, (e) $\log_{10}(f_b/f_d^*)$, (f) $\log_{10}(f_b/f_\rho^*)$ over $\beta_p$ respectively. The starred frequencies are the corresponding frequencies from the modified Taylor hypothesis. The black lines are the linear fitting with least-square method. The averages and standard deviations of each $\beta_p$ bin are plotted as blue lines.\label{fig:5}}
\end{figure}

Figure \ref{fig:5} shows the $\beta_p$ dependence of the ratios. The result is similar to \citet{Chen2014}. $f_b$ locates around $f_d$ ($f_b/f_d \approx 1$) where $\beta_p \ll 1$, while $f_b$ locates around $f_\rho$ ($f_b/f_\rho \approx 1$) where $\beta_p \gg 1$. $f_b$ approaches $f_c$ ($f_b/f_c \approx 1$) for all The modified ratios have the similar trends. The correlation coefficients are shown in Table \ref{tab:2}. As $f_c=(1/f_d+1/f_\rho)^{-1}$, the $f_c$ is close to the smaller of $f_d$ and $f_\rho$. Our result could not distinguish the behavior of the different possibilities.

\section{Conclusion and Discussion}
\label{sec:conclusion}
We have investigated the radial and $\beta_p$ dependence of the observed proton-scale magnetic spectral break frequency $f_b$ in the slow solar wind from 0.17 AU $<r<$ 0.63 AU. Additionally, we have compared the break scale with the spacecraft frequencies corresponding to the cyclotron resonance, $f_c$, proton gyroscale, $f_\rho$, and proton inertial scale, $f_d$ over the range of the heliocentric distance, $r$. The results show that the break frequency follows a power law of $f_b \propto r^{-1.11}$. We find that the break frequency has  mild correlation with all of the three plasma characteristic scales. There is no clearly statistic difference between the result from the plain and the modified Taylor hypothesis. However, $f_b/f_c$ is closest to unity over the full range of distances covered. Nevertheless since the predicted breaks scales are typically only defined to order unity, it is difficult to distinguish them at the moderate values of $\beta_p$ observed by \textit{PSP} to date.

This work provides the first measurement of the radial scaling of the proton-scale break in the slow solar wind in the inner heliosphere down to 0.17 AU. The slow solar wind break manifests a radial dependence similar to the fast wind, with the spectral break occurring around the ion cyclotron resonance scales\citep{Bruno2014}. This suggests that cyclotron resonance may be an important process in the slow solar wind, similar to observations at 1 AU, although the anisotropy of the turbulence complicates a simple picture of parallel-wavenumber cyclotron damping of Alfv\'en waves.

The ratio $f_b/f_c$ approaches unity near the Sun, which may be due to the increased activity of the solar wind plasma close to the Sun generating ion cyclotron waves \citep{Bowen2019}. Regarding that $f_b/f_c$ deviates slightly from unity (less than unity) in the slow solar wind at 1 AU \citep{Woodham2018} and that $f_b/f_c$ increases slightly with decreasing heliocentric distance in the slow solar wind, it seems to be a natural result for $f_b/f_c$ to approach unity near the Sun.

Considering that $f_b$ correlates with all three of $f_c$, $f_d$ and $f_\rho$, we cannot constrain the physical mechanisms which relate to the spectral break. For instance, the observations of \citet{Vech2018} which suggest that magnetic reconnection may disrupt the inertial cascade \citep{Mallet2017} at a disruption scale which has a similar scaling to the cyclotron resonant scale, if proton and electron temperatures are similar. Due to our current lack of electron temperature we have not made any attempt to distinguish the disruption scale.

Near the Sun, the interpretation of the spectral break should be taken carefully. One reason is the failure of the Taylor hypothesis. Our result of the modified Taylor hypothesis from \citet{Klein2015} is only available for the outward-propagating fluctuations in the turbulence dominant with the outward-propagating components. Whether this modification is still available at the future perihelions is still unknown.  Another reason is that the large amplitude fluctuations of magnetic fields and proton bulk velocities are found prevalent near the Sun \citep{Bale2019, Kasper2019}. The generation and the role of these structures in the solar wind turbulence is an open question. In this paper, these fluctuations are treated as a part of the turbulent cascade. The behavior of the spectral break in these structures need a further elucidation. 

As \textit{PSP} descends deeper into the heliosphere we expect to study the break scale where physical scales show better separation in spacecraft frequency. In addition to studying the spectral break, investigation into the dynamics of particles and waves at kinetic scales may constrain the process by which the spectra steepens. The observational studies find that the kinetic fluctuations could be quasi-parallel ion cyclotron waves, quasi-perpendicular kinetic Alfv\'en waves, or the combination of both types at 1 AU \citep{He2011,He2012a, He2012, Salem2012, Klein2014, Zhao2017}. The behavior of the fluctuation near the break scale in the inner heliosphere needs a more comprehensive analysis. As the evidences of the magnetic reconnection and accompanying turbulent enhancement are found in the solar wind \citep{Gosling2004,Phan2006,He2018}, the kinetic-scale fluctuation from the reconnection is another possible explanation to the spectral break. The contribution of the reconnection comparing with other mechanisms requires a quantitative clarification.

\begin{deluxetable*}{ccc}
\tablenum{1}
\tablecaption{The selected fitting frequency interval for the dissipation range\label{tab:1}}
\tablewidth{0pt}
\tablehead{
\colhead{Date} & \colhead{$r$ (AU)} & \colhead{Frequency (Hz)} 
}
\startdata
Mar 10-11 & 0.60-0.63 & 0.8-1.4 \\
Mar 12-15 & 0.54-0.60 & 0.9-1.4 \\
Mar 17-19 & 0.47-0.52 & 0.9-1.5\\
Mar 20-24 & 0.37-0.47 & 1.2-2.2\\
Mar 25-28 & 0.28-0.37 & 1.5-2.5\\
Mar 29-30 & 0.23-0.28 & 1.5-3\\
Mar 31-Apr 5 & 0.17-0.23 & 2-5 
\enddata
\end{deluxetable*}

\begin{deluxetable*}{cccc}
\tablenum{2}
\tablecaption{Summary of the correlation coefficients of various power-law fits}
\tablewidth{0pt}
\label{tab:2}
\tablehead{
\colhead{Parameter 1} & \colhead{Parameter 2} &  \colhead{PCC} 
&\colhead{SCC}}
\startdata
  &$r$&-0.81&-0.84\\
$f_b$&$v_{sw}$ &0.11&0.10\\
&$\beta_{p}$&-0.45&-0.51\\
\cline{1-4}
&$f_c$  & 0.78 & 0.76 \\
$f_b$&$f_d$ &0.70 & 0.64\\
 &$f_\rho$  & 0.69& 0.72\\
\cline{1-4}
 & $f_b/f_c$ & -0.40  & -0.34\\
$r$ & $f_b/f_d$ & -0.61 & -0.63\\
 & $f_b/f_\rho$ & 0.13 & 0.30\\
\cline{1-4}
 & $f_b/f_c^*$ & -0.07  & -0.03\\
$r$ & $f_b/f_d^*$ & -0.40 & -0.39\\
 & $f_b/f_\rho^*$ & 0.32 & 0.47\\
\cline{1-4}
 & $f_b/f_c$ & -0.05 & -0.09 \\
$\beta_p$ & $f_b/f_d$ & -0.55 & -0.59 \\
 & $f_b/f_\rho$ & 0.71 & 0.71 \\
 \cline{1-4}
 & $f_b/f_c^*$ & 0.27 & 0.27 \\
$\beta_p$ & $f_b/f_d^*$ & -0.36 & -0.38 \\
 & $f_b/f_\rho^*$ & 0.82 & 0.82
\enddata
\end{deluxetable*}

\bigbreak

\noindent Acknowledgements:
We thank the referee for helpful comments and the NASA Parker Solar Probe Mission and the FIELDS and SWEAP teams for use of data. D.D. is supported by the China Scholarship Council for his stay at SSL. C.H.K.C. is supported by STFC Ernest Rutherford Fellowship ST/N003748/2. The FIELDS and the SWEAP experiment on the Parker Solar Probe spacecraft was designed and developed under NASA contract NNN06AA01C. D.D. and J.S.H. are also supported by NSFC under 41874200, 41574168, and 41421003. The authors acknowledge the extraordinary contributions of the Parker Solar Probe mission operations and spacecraft engineering teams at the Johns Hopkins University Applied Physics Laboratory. PSP data is available on SPDF (https://cdaweb. sci.gsfc.nasa.gov/index.html/).

\bibliography{main}{}

\begin{thebibliography}{}
\expandafter\ifx\csname natexlab\endcsname\relax\def\natexlab#1{#1}\fi
\providecommand{\url}[1]{\href{#1}{#1}}
\providecommand{\dodoi}[1]{doi:~\href{http://doi.org/#1}{\nolinkurl{#1}}}
\providecommand{\doeprint}[1]{\href{http://ascl.net/#1}{\nolinkurl{http://ascl.net/#1}}}
\providecommand{\doarXiv}[1]{\href{https://arxiv.org/abs/#1}{\nolinkurl{https://arxiv.org/abs/#1}}}

\bibitem[{Alexandrova {et~al.}(2009)Alexandrova, Saur, Lacombe, Mangeney,
  Mitchell, Schwartz, \& Robert}]{Alexandrova2009}
Alexandrova, O., Saur, J., Lacombe, C., {et~al.} 2009, Phys. Rev. Lett., 103,
  165003, \dodoi{10.1103/PhysRevLett.103.165003}

\bibitem[{Bale {et~al.}(2005)Bale, Kellogg, Mozer, Horbury, \&
  Reme}]{Bale2005a}
Bale, S.~D., Kellogg, P.~J., Mozer, F.~S., Horbury, T.~S., \& Reme, H. 2005,
  Physical Review Letters, 94, 215002, \dodoi{10.1103/PhysRevLett.94.215002}

\bibitem[{Bale {et~al.}(2016)Bale, Goetz, Harvey, Turin, Bonnell,
  {Dudok de Wit}, Ergun, MacDowall, Pulupa, Andre, Bolton, Bougeret, Bowen,
  Burgess, Cattell, Chandran, Chaston, Chen, Choi, Connerney, Cranmer,
  Diaz-Aguado, Donakowski, Drake, Farrell, Fergeau, Fermin, Fischer, Fox,
  Glaser, Goldstein, Gordon, Hanson, Harris, Hayes, Hinze, Hollweg, Horbury,
  Howard, Hoxie, Jannet, Karlsson, Kasper, Kellogg, Kien, Klimchuk,
  Krasnoselskikh, Krucker, Lynch, Maksimovic, Malaspina, Marker, Martin,
  Martinez-Oliveros, McCauley, McComas, McDonald, Meyer-Vernet, Moncuquet,
  Monson, Mozer, Murphy, Odom, Oliverson, Olson, Parker, Pankow, Phan,
  Quataert, Quinn, Ruplin, Salem, Seitz, Sheppard, Siy, Stevens, Summers,
  Szabo, Timofeeva, Vaivads, Velli, Yehle, Werthimer, \& Wygant}]{Bale2016}
Bale, S.~D., Goetz, K., Harvey, P.~R., {et~al.} 2016, Space Science Reviews,
  204, 49, \dodoi{10.1007/s11214-016-0244-5}

\bibitem[{Bale {et~al.}(2019)Bale, Badman, Bonnell, Bowen, Burgess, Case,
  Cattell, Chandran, Chaston, Chen, Drake, de~Wit, Eastwood, Ergun, Farrell,
  Fong, Goetz, Goldstein, Goodrich, Harvey, Horbury, Howes, Kasper, Kellogg,
  Klimchuk, Korreck, Krasnoselskikh, Krucker, Laker, Larson, MacDowall,
  Maksimovic, Malaspina, Martinez-Oliveros, McComas, Meyer-Vernet, Moncuquet,
  Mozer, Phan, Pulupa, Raouafi, Salem, Stansby, Stevens, Szabo, Velli, Woolley,
  \& Wygant}]{Bale2019}
Bale, S.~D., Badman, S.~T., Bonnell, J.~W., {et~al.} 2019, Nature,
  \dodoi{10.1038/s41586-019-1818-7}

\bibitem[{Bourouaine {et~al.}(2012)Bourouaine, Alexandrova, Marsch, \&
  Maksimovic}]{Bourouaine2012}
Bourouaine, S., Alexandrova, O., Marsch, E., \& Maksimovic, M. 2012, ApJ, 749,
  102, \dodoi{10.1088/0004-637X/749/2/102}

\bibitem[{Bourouaine \& Chandran(2013)}]{Bourouaine2013}
Bourouaine, S., \& Chandran, B.~D. 2013, The Astrophysical Journal, 774, 96

\bibitem[{Bourouaine \& Perez(2018)}]{Bourouaine2018}
Bourouaine, S., \& Perez, J.~C. 2018, ApJ, 858, L20,
  \dodoi{10.3847/2041-8213/aabccf}

\bibitem[{Bourouaine \& Perez(2019)}]{Bourouaine2019}
---. 2019, ApJ, 879, L16, \dodoi{10.3847/2041-8213/ab288a}

\bibitem[{Bowen(2020, submitted, this volume)}]{Bowen2019}
Bowen, T.~A. 2020, submitted, this volume, \apj

\bibitem[{{Bowen} {et~al.}(2018){Bowen}, {Mallet}, {Bonnell}, \&
  {Bale}}]{Bowen2018}
{Bowen}, T.~A., {Mallet}, A., {Bonnell}, J.~W., \& {Bale}, S.~D. 2018, \apj,
  865, 45, \dodoi{10.3847/1538-4357/aad95b}

\bibitem[{Bruno \& Carbone(2013)}]{Bruno2013}
Bruno, R., \& Carbone, V. 2013, Living Reviews in Solar Physics, 10,
  \dodoi{10.12942/lrsp-2013-2}

\bibitem[{Bruno \& Trenchi(2014{\natexlab{a}})}]{Bruno2014}
Bruno, R., \& Trenchi, L. 2014{\natexlab{a}}, ApJ, 787, L24,
  \dodoi{10.1088/2041-8205/787/2/L24}

\bibitem[{Bruno \& Trenchi(2014{\natexlab{b}})}]{Bruno2014b}
---. 2014{\natexlab{b}}, Astrophysical Journal Letters, 787, 2,
  \dodoi{10.1088/2041-8205/787/2/L24}

\bibitem[{Chandran {et~al.}(2010)Chandran, Li, Rogers, Quataert, \&
  Germaschewski}]{Chandran2010}
Chandran, B.~D., Li, B., Rogers, B.~N., Quataert, E., \& Germaschewski, K.
  2010, ApJ, 720, 503, \dodoi{10.1088/0004-637X/720/1/503}

\bibitem[{Chaspis(2020, to be submitted)}]{Chaspis2019}
Chaspis, A. 2020, to be submitted, \apj

\bibitem[{Chen(2016)}]{Chen2016}
Chen, C. H.~K. 2016, Journal of Plasma Physics, 82, 535820602,
  \dodoi{10.1017/S0022377816001124}

\bibitem[{Chen(2020, submitted, this volume)}]{Chen2019}
---. 2020, submitted, this volume, \apj

\bibitem[{Chen {et~al.}(2010)Chen, Horbury, Schekochihin, Wicks, Alexandrova,
  \& Mitchell}]{Chen2010b}
Chen, C. H.~K., Horbury, T.~S., Schekochihin, A.~A., {et~al.} 2010, Phys. Rev.
  Lett., 104, 255002, \dodoi{10.1103/PhysRevLett.104.255002}

\bibitem[{Chen {et~al.}(2014)Chen, Leung, Boldyrev, Maruca, \& Bale}]{Chen2014}
Chen, C. H.~K., Leung, L., Boldyrev, S., Maruca, B.~A., \& Bale, S.~D. 2014,
  Geophysical Research Letters, 41, 8081, \dodoi{10.1002/2014GL062009}

\bibitem[{Chhiber {et~al.}(2019)Chhiber, Usmanov, Matthaeus, Parashar, \&
  Goldstein}]{Chhiber2019}
Chhiber, R., Usmanov, A.~V., Matthaeus, W.~H., Parashar, T.~N., \& Goldstein,
  M.~L. 2019, The Astrophysical Journal Supplement Series, 242, 12,
  \dodoi{10.3847/1538-4365/ab16d7}

\bibitem[{Cho \& Vishniac(2000)}]{Cho2000}
Cho, J., \& Vishniac, E.~T. 2000, \apj, 539, 273, \dodoi{10.1086/309213}

\bibitem[{Dmitruk {et~al.}(2004)Dmitruk, Matthaeus, \& Seenu}]{Dmitruk2004}
Dmitruk, P., Matthaeus, W.~H., \& Seenu, N. 2004, ApJ, 617, 667,
  \dodoi{10.1086/425301}

\bibitem[{Duan {et~al.}(2018)Duan, He, Pei, Huang, Wu, Verscharen, \&
  Wang}]{Duan2018}
Duan, D., He, J., Pei, Z., {et~al.} 2018, ApJ, 865,
  \dodoi{10.3847/1538-4357/aad9aa}

\bibitem[{Fox {et~al.}(2016)Fox, Velli, Bale, Decker, Driesman, Howard, Kasper,
  Kinnison, Kusterer, Lario, Lockwood, McComas, Raouafi, \& Szabo}]{Fox2016}
Fox, N.~J., Velli, M.~C., Bale, S.~D., {et~al.} 2016, Space Science Reviews,
  204, 7, \dodoi{10.1007/s11214-015-0211-6}

\bibitem[{Franci {et~al.}(2016)Franci, Landi, Matteini, Verdini, \&
  Hellinger}]{Franci2016}
Franci, L., Landi, S., Matteini, L., Verdini, A., \& Hellinger, P. 2016, ApJ,
  833, 91, \dodoi{10.3847/1538-4357/833/1/91}

\bibitem[{Galtier(2006)}]{Galtier2006}
Galtier, S. 2006, Journal of Plasma Physics, 72, 721–769,
  \dodoi{10.1017/S0022377806004521}

\bibitem[{Gosling {et~al.}(2005)Gosling, Skoug, McComas, \&
  Smith}]{Gosling2004}
Gosling, J.~T., Skoug, R.~M., McComas, D.~J., \& Smith, C.~W. 2005, Journal of
  Geophysical Research: Space Physics, 110, \dodoi{10.1029/2004JA010809}

\bibitem[{He {et~al.}(2011)He, Marsch, Tu, Yao, \& Tian}]{He2011}
He, J., Marsch, E., Tu, C., Yao, S., \& Tian, H. 2011, ApJ, 731, 85,
  \dodoi{10.1088/0004-637X/731/2/85}

\bibitem[{He {et~al.}(2013)He, Tu, Marsch, Bourouaine, \& Pei}]{He2013}
He, J., Tu, C., Marsch, E., Bourouaine, S., \& Pei, Z. 2013, ApJ, 773, 72,
  \dodoi{10.1088/0004-637X/773/1/72}

\bibitem[{He {et~al.}(2012{\natexlab{a}})He, Tu, Marsch, \& Yao}]{He2012a}
He, J., Tu, C., Marsch, E., \& Yao, S. 2012{\natexlab{a}}, ApJ, 745, L8,
  \dodoi{10.1088/2041-8205/745/1/L8}

\bibitem[{He {et~al.}(2012{\natexlab{b}})He, Tu, Marsch, \& Yao}]{He2012}
---. 2012{\natexlab{b}}, ApJ, 749, 86, \dodoi{10.1088/0004-637X/749/1/86}

\bibitem[{{He} {et~al.}(2018){He}, {Zhu}, {Chen}, {Salem}, {Stevens}, {Li},
  {Ruan}, {Zhang}, \& {Tu}}]{He2018}
{He}, J., {Zhu}, X., {Chen}, Y., {et~al.} 2018, \apj, 856, 148,
  \dodoi{10.3847/1538-4357/aab3cd}

\bibitem[{Horbury {et~al.}(2008)Horbury, Forman, \& Oughton}]{Horbury2008}
Horbury, T.~S., Forman, M., \& Oughton, S. 2008, Physical Review Letters, 101,
  175005, \dodoi{10.1103/PhysRevLett.101.175005}

\bibitem[{{Howes}(2017)}]{Howes2017Ph}
{Howes}, G.~G. 2017, Physics of Plasmas, 24, 055907, \dodoi{10.1063/1.4983993}

\bibitem[{Howes {et~al.}(2008)Howes, Cowley, Dorland, Hammett, Quataert, \&
  Schekochihin}]{Howes2008}
Howes, G.~G., Cowley, S.~C., Dorland, W., {et~al.} 2008, Journal of Geophysical
  Research: Space Physics, 113, 1, \dodoi{10.1029/2007JA012665}

\bibitem[{Kasper {et~al.}(2016)Kasper, Abiad, Austin, Balat-Pichelin, Bale,
  Belcher, Berg, Bergner, Berthomier, Bookbinder, Brodu, Caldwell, Case,
  Chandran, Cheimets, Cirtain, Cranmer, Curtis, Daigneau, Dalton, Dasgupta,
  DeTomaso, Diaz-Aguado, Djordjevic, Donaskowski, Effinger, Florinski, Fox,
  Freeman, Gallagher, Gary, Gauron, Gates, Goldstein, Golub, Gordon, Gurnee,
  Guth, Halekas, Hatch, Heerikuisen, Ho, Hu, Johnson, Jordan, Korreck, Larson,
  Lazarus, Li, Livi, Ludlam, Maksimovic, McFadden, Marchant, Maruca, McComas,
  Messina, Mercer, Park, Peddie, Pogorelov, Reinhart, Richardson, Robinson,
  Rosen, Skoug, Slagle, Steinberg, Stevens, Szabo, Taylor, Tiu, Turin, Velli,
  Webb, Whittlesey, Wright, Wu, \& Zank}]{Kasper2016}
Kasper, J.~C., Abiad, R., Austin, G., {et~al.} 2016, Space Science Reviews,
  204, 131, \dodoi{10.1007/s11214-015-0206-3}

\bibitem[{Kasper {et~al.}(2019)Kasper, Bale, Belcher, Berthomier, Case,
  Chandran, Curtis, Gallagher, Gary, Golub, Halekas, Ho, Horbury, Hu, Huang,
  Klein, Korreck, Larson, Livi, Maruca, Lavraud, Louarn, Maksimovic,
  Martinovic, McGinnis, Pogorelov, Richardson, Skoug, Steinberg, Stevens,
  Szabo, Velli, Whittlesey, Wright, Zank, MacDowall, McComas, McNutt, Pulupa,
  Raouafi, \& Schwadron}]{Kasper2019}
Kasper, J.~C., Bale, S.~D., Belcher, J.~W., {et~al.} 2019, Nature,
  \dodoi{10.1038/s41586-019-1813-z}

\bibitem[{Kiyani {et~al.}(2015)Kiyani, Osman, \& Chapman}]{Kiyani2015}
Kiyani, K.~H., Osman, K.~T., \& Chapman, S.~C. 2015, Philosophical Transactions
  of the Royal Society A: Mathematical, Physical and Engineering Sciences, 373,
  20140155, \dodoi{10.1098/rsta.2014.0155}

\bibitem[{{Klein} {et~al.}(2014){Klein}, {Howes}, {TenBarge}, \&
  {Podesta}}]{Klein2014}
{Klein}, K.~G., {Howes}, G.~G., {TenBarge}, J.~M., \& {Podesta}, J.~J. 2014,
  \apj, 785, 138, \dodoi{10.1088/0004-637X/785/2/138}

\bibitem[{Klein {et~al.}(2015)Klein, Perez, Verscharen, Mallet, \&
  Chandran}]{Klein2015}
Klein, K.~G., Perez, J.~C., Verscharen, D., Mallet, A., \& Chandran, B.~D.
  2015, Astrophysical Journal Letters, 801, L18,
  \dodoi{10.1088/2041-8205/801/1/L18}

\bibitem[{Leamon {et~al.}(2000)Leamon, Matthaeus, Smith, Zank, Mullan, \&
  Oughton}]{Leamon2000}
Leamon, R.~J., Matthaeus, W.~H., Smith, C.~W., {et~al.} 2000, ApJ, 537, 1054,
  \dodoi{10.1086/309059}

\bibitem[{Leamon {et~al.}(1998)Leamon, Smith, Ness, Matthaeus, \&
  Wong}]{Leamon1998}
Leamon, R.~J., Smith, C.~W., Ness, N.~F., Matthaeus, W.~H., \& Wong, H.~K.
  1998, Journal of Geophysical Research, 103, 4775, \dodoi{10.1029/97JA03394}

\bibitem[{Leamon {et~al.}(1999)Leamon, Smith, Ness, \& Wong}]{Leamon1999a}
Leamon, R.~J., Smith, C.~W., Ness, N.~F., \& Wong, H.~K. 1999, Journal of
  Geophysical Research: Space Physics, 104, 22331, \dodoi{10.1029/1999JA900158}

\bibitem[{{Loureiro} \& {Boldyrev}(2017)}]{Loureiro2017}
{Loureiro}, N.~F., \& {Boldyrev}, S. 2017, \apj, 850, 182,
  \dodoi{10.3847/1538-4357/aa9754}

\bibitem[{Mallet {et~al.}(2017)Mallet, Schekochihin, \& Chandran}]{Mallet2017}
Mallet, A., Schekochihin, A.~A., \& Chandran, B. D.~G. 2017, Journal of Plasma
  Physics, 83, 905830609, \dodoi{10.1017/S0022377817000812}

\bibitem[{Markovskii {et~al.}(2008)Markovskii, Vasquez, \&
  Smith}]{markovskii2008statistical}
Markovskii, S., Vasquez, B.~J., \& Smith, C.~W. 2008, The Astrophysical
  Journal, 675, 1576

\bibitem[{Marsch(2006)}]{Marsch2006}
Marsch, E. 2006, Living Reviews in Solar Physics, 3, 1,
  \dodoi{10.12942/lrsp-2006-1}

\bibitem[{Martinovi{\'c} {et~al.}(2019)Martinovi{\'c}, Klein, \&
  Bourouaine}]{martinovic2019radial}
Martinovi{\'c}, M.~M., Klein, K.~G., \& Bourouaine, S. 2019, arXiv preprint
  arXiv:1905.13355

\bibitem[{Narita {et~al.}(2013)Narita, Glassmeier, Motschmann, \&
  Wilczek}]{Narita2013}
Narita, Y., Glassmeier, K.-H., Motschmann, U., \& Wilczek, M. 2013, Earth,
  Planets and Space, 65, e5, \dodoi{10.5047/eps.2012.12.002}

\bibitem[{Perri {et~al.}(2010)Perri, Carbone, \& Veltri}]{Perri2010}
Perri, S., Carbone, V., \& Veltri, P. 2010, ApJ, 725, L52,
  \dodoi{10.1088/2041-8205/725/1/L52}

\bibitem[{{Peter Gary} \& Smith(2009)}]{PeterGary2009}
{Peter Gary}, S., \& Smith, C.~W. 2009, Journal of Geophysical Research: Space
  Physics, 114, 1, \dodoi{10.1029/2009JA014525}

\bibitem[{{Phan} {et~al.}(2006){Phan}, {Gosling}, {Davis}, {Skoug},
  {{\O}ieroset}, {Lin}, {Lepping}, {McComas}, {Smith}, {Reme}, \&
  {Balogh}}]{Phan2006}
{Phan}, T.~D., {Gosling}, J.~T., {Davis}, M.~S., {et~al.} 2006, \nat, 439, 175,
  \dodoi{10.1038/nature04393}

\bibitem[{{Salem} {et~al.}(2012){Salem}, {Howes}, {Sundkvist}, {Bale},
  {Chaston}, {Chen}, \& {Mozer}}]{Salem2012}
{Salem}, C.~S., {Howes}, G.~G., {Sundkvist}, D., {et~al.} 2012, \apjl, 745, L9,
  \dodoi{10.1088/2041-8205/745/1/L9}

\bibitem[{Schekochihin {et~al.}(2009)Schekochihin, Cowley, Dorland, Hammett,
  Howes, Quataert, \& Tatsuno}]{Schekochihin2009}
Schekochihin, A.~A., Cowley, S.~C., Dorland, W., {et~al.} 2009, The
  Astrophysical Journal Supplement Series, 182, 310,
  \dodoi{10.1088/0067-0049/182/1/310}

\bibitem[{Stawicki {et~al.}(2001)Stawicki, Gary, \& Li}]{Stawicki2001}
Stawicki, O., Gary, S.~P., \& Li, H. 2001, Journal of Geophysical Research:
  Space Physics, 106, 8273, \dodoi{10.1029/2000ja000446}

\bibitem[{Vasquez {et~al.}(2007)Vasquez, Abramenko, Haggerty, \&
  Smith}]{Vasquez2007}
Vasquez, B.~J., Abramenko, V.~I., Haggerty, D.~K., \& Smith, C.~W. 2007,
  Journal of Geophysical Research: Space Physics, 112, n/a,
  \dodoi{10.1029/2007JA012504}

\bibitem[{Vech {et~al.}(2017)Vech, Klein, \& Kasper}]{Vech2017}
Vech, D., Klein, K.~G., \& Kasper, J.~C. 2017, The Astrophysical Journal
  Letters, 850, L11, \dodoi{10.3847/2041-8213/aa9887}

\bibitem[{Vech {et~al.}(2018)Vech, Mallet, Klein, \& Kasper}]{Vech2018}
Vech, D., Mallet, A., Klein, K.~G., \& Kasper, J.~C. 2018, The Astrophysical
  Journal Letters, 855, L27, \dodoi{10.3847/2041-8213/aab351}

\bibitem[{Verscharen {et~al.}(2019)Verscharen, Klein, \&
  Maruca}]{verscharen2019multi}
Verscharen, D., Klein, K.~G., \& Maruca, B.~A. 2019, arXiv preprint
  arXiv:1902.03448

\bibitem[{Wang {et~al.}(2018)Wang, Tu, He, \& Wang}]{Wang2018}
Wang, X., Tu, C.~Y., He, J.~S., \& Wang, L.~H. 2018, Journal of Geophysical
  Research: Space Physics, 123, 68, \dodoi{10.1002/2017JA024813}

\bibitem[{{Woodham} {et~al.}(2018){Woodham}, {Wicks}, {Verscharen}, \&
  {Owen}}]{Woodham2018}
{Woodham}, L.~D., {Wicks}, R.~T., {Verscharen}, D., \& {Owen}, C.~J. 2018,
  \apj, 856, 49, \dodoi{10.3847/1538-4357/aab03d}

\bibitem[{Yan {et~al.}(2016)Yan, He, Zhang, Tu, Marsch, Chen, Wang, Wang, \&
  Wicks}]{Yan2016}
Yan, L., He, J., Zhang, L., {et~al.} 2016, ApJ, 816, L24,
  \dodoi{10.3847/2041-8205/816/2/L24}

\bibitem[{{Zhao} {et~al.}(2017){Zhao}, {Chu}, {Lin}, {Yang}, {Feng}, {Wu}, \&
  {Liu}}]{Zhao2017}
{Zhao}, G.~Q., {Chu}, Y.~H., {Lin}, P.~H., {et~al.} 2017, Journal of
  Geophysical Research (Space Physics), 122, 4879, \dodoi{10.1002/2017JA024119}

\end{thebibliography}
\bibliographystyle{aasjournal}



\end{document}